\documentclass[conference]{IEEEtran}
\IEEEoverridecommandlockouts
\usepackage[utf8]{inputenc}

\usepackage{amsmath}
\usepackage{amsfonts}
\usepackage{amssymb}
\usepackage{graphicx}
\graphicspath{ {./images/} }
\usepackage{tikz}
\usetikzlibrary{positioning}
\usetikzlibrary{shapes.geometric,arrows}
\usetikzlibrary{calc,positioning, fit}
\usepackage{caption}
\usepackage{float}
\usepackage{subcaption}
\usepackage[export]{adjustbox}
\usepackage{bm}
\usepackage{filecontents,lipsum}
\usepackage[noadjust]{cite}
\usepackage{adjustbox}

\begin{document}

\title{A Periodogram-Based Method to Identify Forced and Natural Oscillations Using PMUs
\thanks{This work is supported by the Fonds de Recherche du Québec - Nature et technologies under Grant FRQ-NT NC-253053 and by Mitacs Globalink Research Internship program.}}

\author{\IEEEauthorblockN{Qinye Tang}
\IEEEauthorblockA{
Zhejiang University\\
Hangzhou, China\\
Email:qtang@zju.edu.cn }
\and
\IEEEauthorblockN{Xiaozhe Wang}
\IEEEauthorblockA{McGill University\\
Montreal, QC Canada\\
Email: xiaozhe.wang2@mcgill.ca}
}

\maketitle

\begin{abstract}
Sustained oscillations in power systems are dangerous. There are various mechanisms, for instance, limit cycle and forced oscillation, that may lead to sustained oscillations, which nevertheless are hard to differentiate. In this paper, a novel periodogram-based method to distinguish different oscillation mechanism is proposed, which can quantitatively extract essential signatures of different mechanisms from power spectral density. Numerical study shows that the proposed method can accurately distinguish different oscillation mechanisms even when the forced oscillation frequency is close to the natural frequency. 

\end{abstract}

\IEEEpeerreviewmaketitle

\tikzstyle{rect}=[draw,rectangle,fill=red!2,text width=1.75cm,text centered,minimum height=0.8cm]
\tikzstyle{rect_for_op}=[draw,rectangle,fill=blue!10,text width=1.75cm,text centered,minimum height=0.8cm]
\tikzstyle{rect_for_decision}=[draw,rectangle,fill=yellow!10,text width=1.75cm,text centered,minimum height=1cm]
\tikzstyle{diam}=[draw,diamond,fill=green!5,text width=1.25cm,text badly centered,inner sep=0pt,minimum height = 0.5cm]
\tikzstyle{ellipse}=[draw,Ellipse,fill=blue!30,text width=3cm,text centered,minimum height=1cm]
\tikzstyle{line}=[draw,-latex']
\maketitle

\section{Introduction}
One of the major threats to the security and stability of power systems is sustained oscillations \cite{wang2016data}\cite{ghorbaniparvar2017survey}, 
which can damage equipment, degrade the power quality and increase the risks of cascading power failure. Oscillations in power systems can be categorized into natural oscillations and forced oscillations. Natural oscillations often result from inherent interaction among dynamic devices, which typically can be mitigated by power system stabilizer (PSS), intertie line controls, etc.  In contrast, forced oscillations refer to system responses to an external periodic perturbation \cite{ghorbaniparvar2017survey}, which cannot be depressed by PSS. The only effective measure is to locate and remove the external driving source.

In previous literature, various methods have been proposed to identify natural oscillations including prony analysis \cite{hauer1990initial}, Frequency Domain Decomposition Analysis \cite{liu2008oscillation},  Subspace Identification \cite{sarmadi2014electromechanical}, Robust RLS method \cite{zhou2007robust}, etc. However, these methods may not effectively identify forced oscillations. To distinguish different oscillation mechanisms based on PMU data, Liu et al. in \cite{liu2014active} used support vector machines (SVM) to analyze different features of the two oscillation mechanisms. However, this method's accuracy relies directly on the envelop size of oscillation. Xie et al. in \cite{xie2015distinguishing} proposed a distinguishing method based on spectrum, yet its performance may degrade when the external force frequency is close to system's natural frequency. Ghorbaniparvar et al. in \cite{ghorbaniparvar2017forecasting} proposed a residual spectral analysis method leveraging on different residual spectral properties of the two oscillation mechanisms. However, a group of forecasting models needs to be calculated, which may hamper its real-time implementation. Wang et al. in \cite{wang2016data}, used a periodogram-based method to distinguish limit cycle oscillation from forced oscillation, but its performance is not guaranteed if the bandwidth of the power spectral density (PSD) is blurred by strong noise or weak external force.  

In this paper,  a novel periodogram-based method to differentiate forced oscillation from natural oscillation is proposed. It will be shown that, mathematically, the limit cycle and the forced oscillation can be described by the same governing stochastic differential equation yet with different parameter values, based on which a PSD-based classifier can be used to distinguish different mechanisms. The proposed method can quantitatively extract signatures of different mechanisms from PSD and provide accurate classification results even when the forced oscillation frequency is close to the natural oscillation frequency.

\section{mathematical models for sustained oscillations in power systems}
The power system dynamic model can be described as:
\small
\begin{eqnarray}
    \dot{\bm{x}} &=& \bm{f}(\bm{x}, \bm{y})\nonumber\\
\bm{0} &=& \bm{g}(\bm{x}, \bm{y}, \bm{u})\label{eq:powermodel}
\end{eqnarray}
\normalsize
where $\bm{x}\in\mathbb{R}^{n_x}$ is the corresponding state variables (generator rotor angles, rotor speeds, etc), $\bm{y}\in\mathbb{R}^{n_y}$ is
algebraic variables (bus voltages, bus angles, etc), and $\bm{u}$ is the vector describing stochastic behaviors in real-world power systems. 

Our main interest lies in the stochastic perturbations like load variations and renewable generations, which can be modeled as a vectorized Ornstein-Uhlenbeck process $\dot{\bm{u}} = -C\bm{u} + \sigma\bm{\xi}$, which is stationary, Gaussian and Markovian \cite{wang2016data}.
$C$ is a diagonal matrix related to the reversing times of the process; $\bm{\xi}$ is a vector of independent standard Gaussian white noise; $\sigma$ the intensity of noise.

Replacing $\bm{y}$ by $\bm{x}$ and $\bm{u}$ using implicit function theorem and linearizing (\ref{eq:powermodel}) around its steady state, we have \cite{wang2015long}:
\small
\begin{equation}
\dot{\bm{q}} = A \bm{q} + \sigma B\bm{\xi}\label{eq:powersystemeqn}
\end{equation}
\normalsize
where $\bm{q} = \begin{bmatrix}\delta x,\delta u\end{bmatrix}$ , $A = \begin{bmatrix}f_x-f_y g_y^{-1}g_x&-f_y g_y^{-1}g_u\\0&-C\end{bmatrix}$,
$B = \begin{bmatrix}0,I_{n_u}\end{bmatrix}^T$.
In the rest of the paper, we focus on the stochastic dynamic model described above, where $\bm{q}$ is a vector Ornstein-Uhlenbeck
process.

Sustained oscillations of power systems can be caused by different mechanisms and thus are represented by different mathematical models. In this work, we are interested in identifying the forced oscillation caused by external force and the limit cycle governed by system inherent dynamics.

\subsection{Forced Oscillation}

 The normal form of the forced oscillation can be described as:
 \begin{equation}
 \dot{\bm{q}}=\begin{bmatrix}\gamma&-\omega_0\\\omega_0&\gamma\end{bmatrix}\bm{q}+F\begin{bmatrix}\cos{\Omega t}\\\sin{\Omega t}\end{bmatrix}+\sigma\begin{bmatrix}\xi_1\\\xi_2\end{bmatrix}
 \end{equation}
 where  $\bm{q}\in\mathbb{R}^2$, $\gamma<0$, and $\xi_1(t)$, $\xi_2(t)$ are independent standard Gaussian white noises. $F$ denotes the external driving force acted to the system with a frequency $\Omega$.

We want to find the closed-form solution of $\bm{q}(t)$ (particularly the first component), from which a nice structure of forced oscillation can be derived. Denote
$M\triangleq \begin{bmatrix}\gamma&-\omega_0\\\omega_0&\gamma\end{bmatrix}$,
then the solution $\bm{q}(t)= \bm{q_c}(t) + \bm{q_{ou}}(t)=\begin{bmatrix}x_c(t)+x_{ou}(t),y_c(t)+y_{ou}(t)\end{bmatrix}^T$ can be found from the control process: 
$\dot{\bm{q_c}}=M\bm{q_c}+F\begin{bmatrix}\cos{\Omega t},\sin{\Omega t}\end{bmatrix}^T$, 
and the Ornstein-Uhlenbeck process:
$\dot{\bm{q}}_{\bm{ou}}=M\bm{q_{ou}}+\sigma\begin{bmatrix}\xi_1,\xi_2\end{bmatrix}^T$. 
The control process 
can be easily solved by evaluating the matrix exponential:
\begin{equation}
 x_c=\begin{bmatrix}1&0\end{bmatrix}\int_0^t e^{(t-s)M}F\begin{bmatrix}\cos{\Omega s}\\\sin{\Omega s}\end{bmatrix}ds=\frac{F\cos{(\Omega t)}}{\sqrt{\gamma^2+(\Omega-\omega_0)^2}}
\end{equation}
Similarly, $x_{ou}$ can be solved by evaluating the It\^{o} integral:
\begin{equation}
    x_{ou}=\begin{bmatrix}1&0\end{bmatrix}\int_0^t e^{(t-s)M}\sigma\begin{bmatrix}dB_s^{(1)}\\dB_s^{(2)}\end{bmatrix}ds=\sigma\int_0^te^{-\gamma(t-s)}dB_s
\end{equation}
where the Brownian terms $dB_s^{(1)}$ and $dB_s^{(2)}$ are independent. 
Therefore, the first component of $\bm{q}$ is of the form:
$x(t)=x_c(t)+x_{ou}(t)$. 
By taking time derivatives on both sides and collecting terms, we obtain:
\begin{equation}
\dot{x}(t)=-\gamma(x(t)-\frac{F}{\gamma}\sqrt{\frac{\gamma^2+\Omega^2}{\gamma^2+(\Omega-\omega_0)^2}}\cos{(\Omega t}))+\sigma\xi(t)\label{eq:forcedsoln}
\end{equation}
where $\xi(t)$ is standard Gaussian white noise (see detailed derivation in Appendix \ref{appendix-1}). 

In other words,  (\ref{eq:forcedsoln}) can be represented as:
\begin{equation}
\dot{x}(t)=-\beta(x(t)-Q(t))
+\sigma\xi(t) \label{eq:Lforced}
\end{equation}
where $\beta=\gamma$ and $Q(t)
=\frac{F}{\gamma}\sqrt{\frac{\gamma^2+\Omega^2}{\gamma^2+(\Omega-\omega_0)^2}}\cos{(\Omega t})$. 

In the following section, we'll show that incidentally the limit cycle can also be represented in the form of (\ref{eq:Lforced}), yet with the parameters $\lambda, \alpha, \sigma_1$ locating in different ranges. Leveraging on this difference, we can develop a data-driven diagnostic to tell the two oscillations apart from each other.  

\subsection{Limit Cycle}
 Stable limit cycle emerging from supercritical Hopf bifurcation is another common power system oscillation mechanism. It's usually regarded as an early warning sign of voltage collapse, since the Hopf bifurcation usually precedes the saddle-node bifurcation which leads to final voltage collapse \cite{wang2016data}.
The normal form of the oscillation stochastic system near Hopf bifurcation can be described as:
 \begin{equation}
  \dot{\bm{q}} = \begin{bmatrix}\gamma&-\omega_h\\ \omega_h&\gamma\end{bmatrix}\bm{q}-|\bm{q}|^2\bm{q}+\sigma\begin{bmatrix}\xi_1\\\xi_2\end{bmatrix}
\end{equation}
where  $\bm{q}\in\mathbb{R}^2$, $\gamma>0$, $\xi_1(t)$, $\xi_2(t)$ are independent standard Gaussian white noises.

As discussed in \cite{wang2016data}, near the Hopf bifurcating point, the amplitude of the limit cycle grows with $\sqrt{\gamma}$, and the angular frequency is approximately $\omega_h$. The solution can be approximated as:
\begin{equation}
\bm{q}=\left(\begin{array}{c}x(t)\\y(t)\end{array}\right)\approx\left(\begin{array}{c}(\sqrt{\gamma}+p(t))\cos(\phi(t))\\(\sqrt{\gamma}+p(t))\sin(\phi(t))\end{array}\right)\triangleq\left(\begin{array}{c}N(t)\\N^\prime(t)\end{array}\right)
\end{equation}
where $\dot{\phi}=\omega_h +\frac{\sigma}{\sqrt{\gamma}}\xi_{\phi}$ and $\dot{p}=-2\gamma p+\sigma\xi_{p}$ is an Ornstein-Uhlenbeck process independent of $\phi$, which is a Brownian motion with deterministic drift \cite{louca2015stable}. 

We claim that $x(t)$ can be described by the following dynamic equation:
\begin{equation}
\dot{x}(t)=-\beta(x(t)-N(t))\label{eq:LLC}
\end{equation}
where 
$\beta \rightarrow \infty$. To see this, taking Laplace transform on both sides of (\ref{eq:LLC}), we have
$\hat{x}(s)=\frac{\beta}{\beta+s}\hat{N}(s)+\frac{x(0)}{\beta+s}$.
Hence by the uniqueness of inverse Laplace transform we obtain that
$\lim\limits_{\beta\to\infty}x(t) = \lim\limits_{\beta\to\infty}\mathcal{L}^{-1}\{\hat{x}(s)\}(t) = N(t)$.



\subsection{The Same Governing Equation Yet Different Parameters}

Comparing (\ref{eq:Lforced}) and (\ref{eq:LLC}), it is easy to see that they can be described by the same stochastic differential equation:
\begin{equation}
\dot{x}(t)=-\beta(x(t)-Q(t;\lambda,\alpha,\sigma_1))
+\sigma_2\xi(t) \label{eq:L}
\end{equation}
where 
\small
\begin{equation}
    Q(t;\lambda,\omega,\sigma_1)\approx(\lambda+\sigma_1\int_{0}^{t}e^{-2\lambda^2(t-s)}dW_s)\cos{(\omega t+\frac{\sigma_1}{\lambda}B_t)}\nonumber
\end{equation}
\normalsize
However, the two oscillation mechanisms have different parameter values as summarized in Table \ref{tb:parametercomp}. We intend to utilize this structure that is able to simultaneously describe the two oscillation mechanisms to develop a data-driven method to distinguish the two oscillation mechanisms.
\begin{table}[!ht]
\centering
\begin{tabular}{cccccc}
\hline
&$\beta$&$\lambda$&$\alpha$&$\sigma_1$&$\sigma_2$\\
\hline
forced oscillation &$\gamma$&$ \frac{F}{\gamma}\sqrt{\frac{\gamma^2+\Omega^2}{\gamma^2+(\Omega-\omega_0)^2}}$&
$\Omega$&0&$\sigma$\\
limit cycle & $\infty$&$\sqrt{\gamma}$&$\omega_h$&$\sigma$&$0$\\
\hline 
\end{tabular}
\caption{The parameter values for the two mechanisms}\label{tb:parametercomp}
\vspace{-20pt}
\end{table}
\\

\section{PSD-based Technique to Distinguish Two Oscillation Mechanisms}

Inspired by the work \cite{louca2014distinguishing}, we consider the power spectral density of the dynamical system (\ref{eq:L}), from which a classifier can be readily proposed to distinguish the oscillation mechanisms. In addition, an optimization technique is exploited to to estimate parameters presented in Table \ref{tb:parametercomp}.

\subsection{The Power Spectral Density}

The power spectral density (PSD) of the linear dynamic system (\ref{eq:L}) takes the following form \cite{gardiner1985handbook}:

\begin{equation}
PS[X](\omega)=\frac{\beta^2}{\beta^2+\omega^2}PS[Q](\omega)+\frac{\sigma_2^2}{\beta^2+\omega^2}\label{eq:PSDX}
\end{equation}
where the PSD of process $Q(t;\lambda,\alpha,\sigma_1)$ can be approximated by: 
\begin{equation}
PS[Q](\omega) \approx 
\lambda^2 F(
\frac{\sigma_1^2}{\lambda^2},\alpha,\omega)+
\frac{\sigma_1^2}{4\lambda^2}F(
\frac{\sigma_1^2}{\lambda^2}+4\lambda^2,\alpha,\omega)
\end{equation}
The detailed expression of $F$ is:
\begin{equation}
F(s,\alpha,\omega)=2s\frac{4(\alpha^2+\omega^2)+s^2}{[4(\alpha-\omega)^2+s^2][4(\alpha+\omega)^2+s^2]}\nonumber
\end{equation}

\subsection{A PSD-Based Classifier}
The general perception about the distinction between the limit cycle and the forced oscillation lies in the temporal decoherence \cite{louca2014distinguishing}\cite{demir2000phase}. The forced oscillation is characterized by a long-term temporal coherence because of the periodic driving, in contrast to intrinsic limit cycle, which has increasing spectral bandwidth due to temporal decoherence.

Particularly, the rightmost term in (\ref{eq:PSDX}) describes the effect of the non-decohering perturbation on the periodic solution $Q(t;\lambda,\alpha,\sigma_1)$. Ideally, the rightmost term in (\ref{eq:PSDX}) should be zero for limit cycle ($\sigma_2=0$) and far from zero for forced oscillation ($\sigma_2=\sigma>0$). In light of this, we propose the following classier: 
\begin{equation}
V_{PS}:=
\frac{\frac{\sigma_2^2}{\beta^2}}{PS[x](0)}
=
\frac{\frac{\sigma_2^2}{\beta^2}}{PS[Q(t;\lambda,\alpha,\sigma_1)](0)+\frac{\sigma_2^2}{\beta^2}}
\nonumber
\end{equation}
which relates the power due to non-decohering perturbations to the total power of DC term (i.e, the process power at zero frequency). 
It is straightforward from its form that  $V_{PS}\in[0,1]$. If the periodic time series data originates from a forced oscillation, $\sigma_2>0$ while $PS[Q(t;\lambda,\alpha,0)](0)=0$, indicating that $V_{PS} $ should be close to 1. In contrast, if the cyclic data results from a limit cycle, $\sigma_2=0$ and $\beta=\infty$ ideally, thus $V_{PS}$ should be very close to 0. 

It is worth mentioning that the value of classifier $V_{PS}$ will not be dramatically influenced if the external force frequency $\Omega$ is close to system's mode frequency $\omega_0$, since the only affected parameter is $\lambda$ from Table \ref{tb:parametercomp}, which nevertheless will not greatly affect $V_{PS}$. Such result is also validated in the numerical study in Section \ref{sectionsimulation}.

\subsection{A Data-Driven Method to Distinguish Different Oscillation Mechanisms}

We have shown that the dynamics of two oscillation mechanisms, namely, the forced oscillation and the limit cycle, can be described with the same governing equation (\ref{eq:L}) with different parameter values. In addition, there is a classifier $V_{ps}$ that possesses a higher value ($\approx 1$) in the case of forced oscillation and a lower value ($\approx 0$) for limit cycle. 

We intend to leverage the above analytical results to develop a diagnostic method to distinguish the two oscillation mechanisms.  
Given PMU data with oscillations, we will first solve two constrained optimization problems to fit the parameters to each oscillation mechanism. The one with less loss function value will be selected for calculating the classifier $V_{PS}$, based on which the oscillation mechanism can be determined.   

More specifically, given a  pair of oscillator's time series $\{x_n\}_{n=1}^N$, $\{y_n\}_{n=1}^N$ and sampling period $\Delta t$, 
we estimate the sample PSD by:
\begin{equation}
PS[x]_{estimated}(\omega) = \frac{\Delta t}{N}|\sum_{n=1}^Nx_ne^{-i\omega n\Delta t}|^2
\end{equation}


Next, we estimate two sets of the five parameters listed in Table \ref{tb:parametercomp} by fitting the estimated sampe PSD. Particularly, two constrained optimization problems are solved. First,  a limit cycle is assumed so that we estimate the parameters by solving:
\small{
\vspace{-10pt}
\begin{align*}
    \min\limits_{\lambda,\alpha,\sigma_1,\beta,\sigma_2}\int_0^{\omega_c}[PS[x]_{estimated}(\omega)&-PS[x(t;\lambda,\alpha,\sigma_1,\beta,\sigma_2)](\omega)]^2d\omega \nonumber\\
    s.t.\qquad\sigma_2&<\epsilon, \quad
    \beta>M \nonumber
\vspace{-10pt}
\end{align*}
}
\normalsize
where $\omega_c$ is the upper bound of the fitting interval; 
$\epsilon$ and 
$M$ are two bounds set to refine the searching space of the parameters according to Table \ref{tb:parametercomp}. 

Second, a forced oscillation is assumed so that we estimate the parameters by solving the second constrained optimization:\\
\small
\begin{align*}
\vspace{-10pt}
    \min\limits_{\lambda,\alpha,\sigma_1,\beta,\sigma_2}\int_0^{\omega_c}[PS[x]_{estimated}(\omega)&-PS[x(t;\lambda,\alpha,\sigma_1,\beta,\sigma_2)](\omega)]^2d\omega\nonumber\\
    s.t.\qquad\sigma_1<\epsilon, & \quad \frac{\sigma_2}{\beta}=PSD[x]_{estimated}(0)\nonumber
\vspace{-10pt}
\end{align*}
\normalsize
where 
$\epsilon$ is an upper bound for $\sigma_1$ based on Table \ref{tb:parametercomp} and the second constraint is applied to guarantee the fitting performance of the DC term. 

It is worth mentioning that it is more efficient to solve two constrained optimization problems for each oscillation mechanism compared to solving one global optimization, since the dimension of the searching space is greatly reduced. 

Once we get two sets of the estimated parameters from the above constrained optimization problems,  
we choose the set of parameters which has less objective function value, i.e., the one that fits the empirical PSD better, to calculate the statistic $V_{PS}$. 
We classify the oscillation to be a forced oscillation if $V_{PS}>0.5$ or to be a limit cycle if $V_{PS}\leq 0.5$.

Note that we do not make the classification decision directly based on the optimization results to avoid potential incorrect results due to overfitting. $V_{PS}$ on the other hand can effectively identify the different signatures of the two oscillation mechanisms around zero frequency and around the peak. 
Thus, combining the analysis on the whole interval $[0 \mbox{ } \omega_c]$  with a particular focus around zero frequency and peak frequency may provide more reliable and accurate identification results.  

\begin{figure}[!ht]
\centering
\scalebox{0.7}{
\small
\begin{tikzpicture}[node distance=17.5mm,auto]
\node[rect,rounded corners](start){PMU Data with Oscillations};
\node[rect,below = 0.25cm of start,label=left:Step1:](PSD){Calculate Sample PSD};
\node[rect,below = 0.5cm of PSD](opt){Optimization using constraints of:};
\node[rect_for_op,rounded corners,below left = 0.5cm and 1mm of opt](LC){Limit Cycle};
\node[rect_for_op,rounded corners,below right = 0.5cm and 1mm of opt](F){Forced Osc};
\node[rect,below = 1.25cm of opt](loss){Calculate both loss function values};
\node[diam,below = 3.25cm of opt](compare){Is forced Osc loss smaller?};
\node[rect_for_op,rounded corners,below left = 0.75cm and 0.5cm of compare](V LC){Calculate $V_{PS}$ using LC fitting parameters};
\node[rect_for_op,rounded corners,below right = 0.75cm and 0.5cm of compare](V F){Calculate $V_{PS}$ using forced Osc fitting parameters};
\node[diam,below = 2cm of compare](final){$V_{PS}>0.5$?};
\node[rect_for_decision,below left = 0.75cm and 0.5cm of final](decision LC){It's limit cycle oscillation!};
\node[rect_for_decision,below right = 0.75cm and 0.5cm of final](decision F){It's forced oscillation!};

\node[draw,black,label=left:Step2:,inner sep=1mm,fit=(opt) (F) (loss) (LC)] {};
\node[draw,black,label=left:Step3:,inner sep=1mm,fit=(V LC)(V LC)(V F)(V F)]{};
\node[draw,black,label=left:Step4:,inner sep=1mm,fit=(decision LC)(decision F)]{};
\path[line](start)--(PSD);
\path[line](PSD)--(opt);
\path[line](opt)--node[left,near start]{}(LC);
\path[line](opt)--node[right,near start]{}(F);
\path[line](LC)--node[right,near start]{}(loss);
\path[line](F)--node[left,near start]{}(loss);
\path[line](loss)--node[below,near start]{}(compare);
\path[line](compare)--node[left,near start]{No}(V LC);
\path[line](compare)--node[right,near start]{Yes}(V F);
\path[line](V LC)--node[right,near start]{}(final);
\path[line](V F)--node[left,near start]{}(final);
\path[line](final)--node[left,near start]{No}(decision LC);
\path[line](final)--node[right,near start]{Yes}(decision F);
\end{tikzpicture}
}
\caption{The Flowchart of the Proposed Method}\label{fig:flowchart}
\vspace{-5pt}
\end{figure}
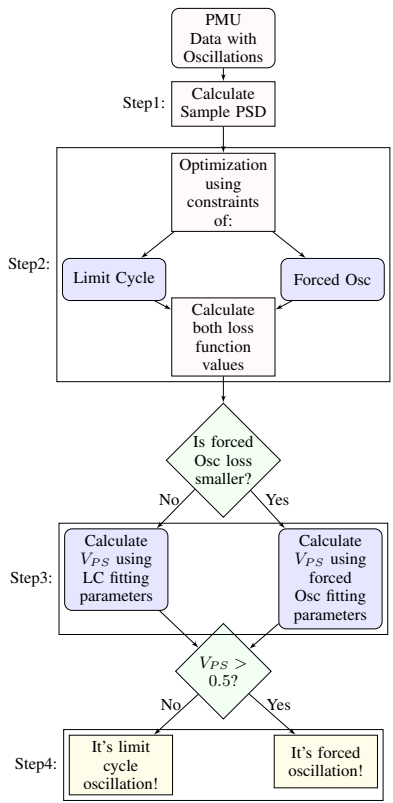

To summarize, assuming that other possibilities like quasi-periodic chaos and weakly-damped oscillation have been ruled out (e.g., via kurtosis \cite{wang2016data}), a flowchart describing the detailed steps of the proposed method to distinguish the limit cycle and the forced oscillation is presented in Fig. \ref{fig:flowchart}.

\section{Simulation Studies} \label{sectionsimulation}
\begin{figure}[!ht]
 \vspace{-5pt}
\centering
\begin{subfigure}{.5\linewidth}
\includegraphics[width=1.6in]{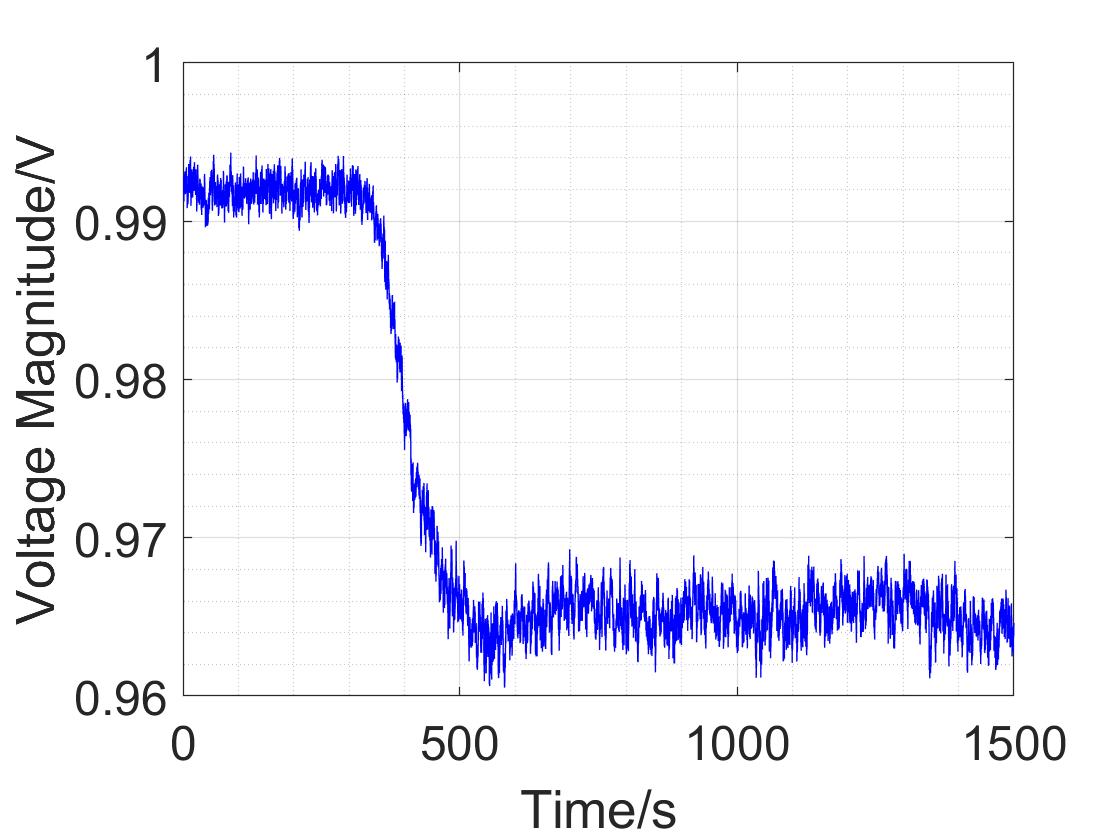}
\caption{Case A Voltage Magnitude}\label{timeseriesA}
\end{subfigure}%
\begin{subfigure}{.5\linewidth}
\includegraphics[width=1.6in]{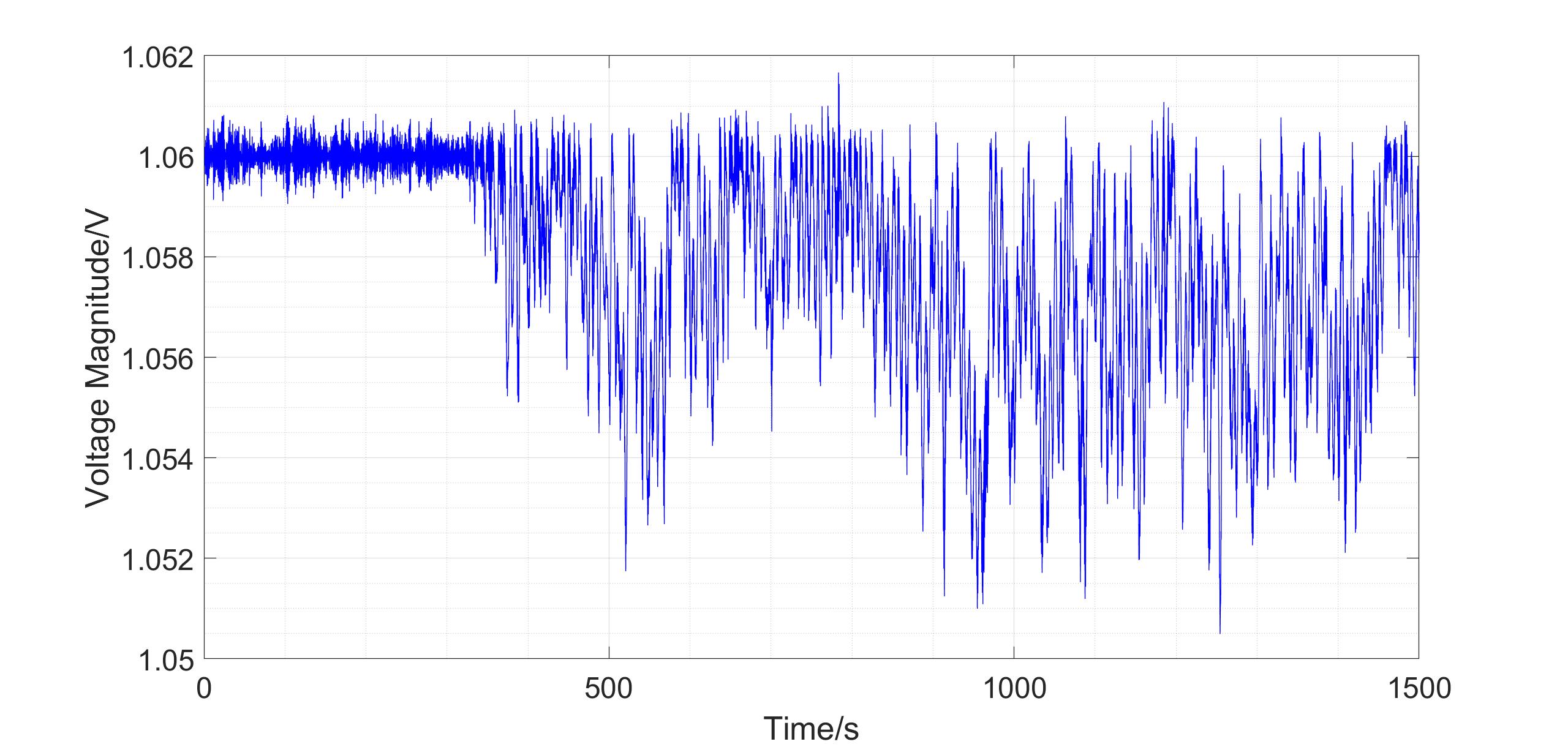}
\caption{Case B Voltage Magnitude}\label{timeseriesB}
\end{subfigure}%
\caption{Evolution of Bus 11's Voltage Magnitude in Two Cases}\label{timeseries}
\end{figure}

Fig. \ref{timeseries} shows two different oscillation scenarios in IEEE 14-bus systems, from which the exact mechanisms can  
be hardly distinguished.
In this section, we test the performance of the  proposed method in identifying the oscillation mechanism of each scenario.  
The emulated PMU data is obtained from the time-domain simulation of the 14-bus system. Stochastic load fluctuations are considered, which follow Orstein-Uhlenbeck process with $\sigma=0.01$ in (\ref{eq:powersystemeqn}). Besides, we use exponential recovery loads to model load dynamics, the model of which can be described as:
\small
\begin{eqnarray}
P&=&kP_0(\frac{V}{V_0})^\alpha\\
Q&=&kQ_0(\frac{V}{V_0})^\beta
\end{eqnarray}
\normalsize
where $k$ is a dimensionless demand variable, $V_0$ is the reference voltage, and $\alpha$ and $\beta$ are active and reactive power exponents depending on the type of load. All simulations were conducted in PSAT-2.1.8\cite{milano2008power}. The parameter values of the test systems are available on: {http://github.com/xiaozhew/Test-Systems-}. 

\begin{figure}[!ht]
\vspace{-15pt}
\centering
\begin{subfigure}{.5\linewidth}
\includegraphics[width=1.8in]{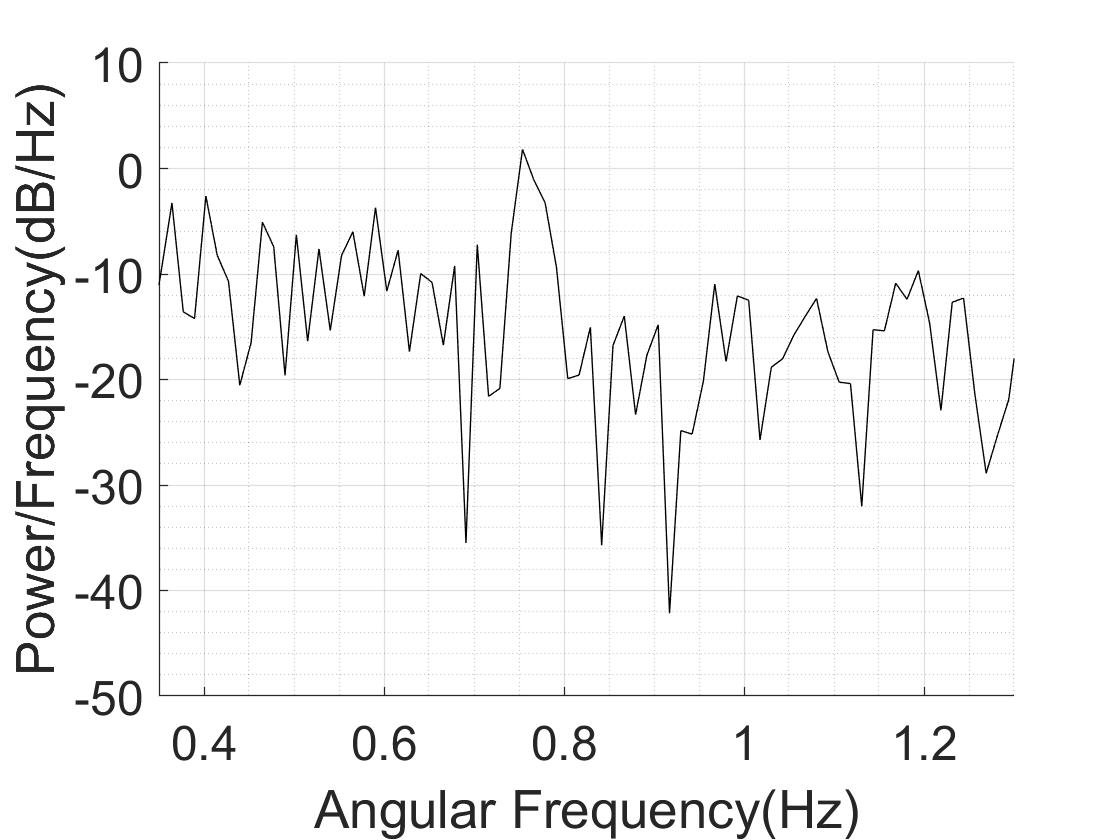}
\caption{Case A PSD}\label{PSDestA}
\end{subfigure}%
\begin{subfigure}{.5\linewidth}
\includegraphics[width=1.8in]{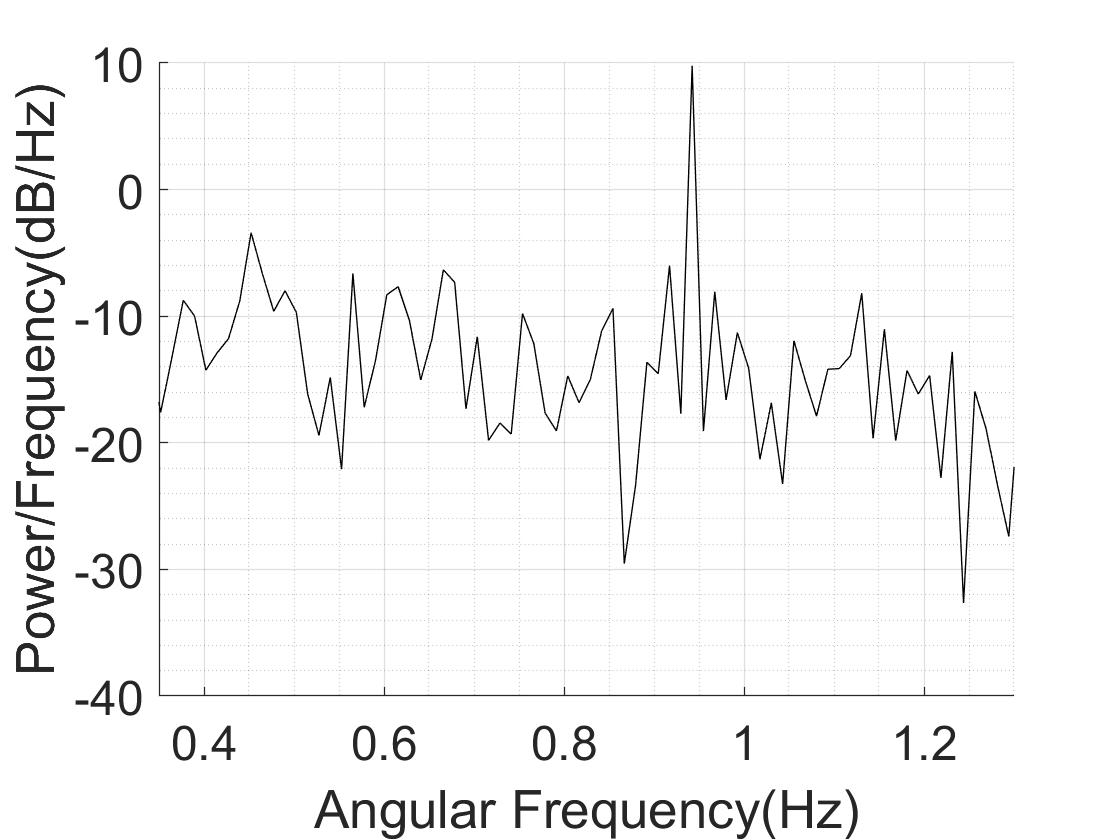}
\caption{Case B PSD}\label{PSDestB}
\end{subfigure}%
\caption{The estimated PSD of Both Cases}\label{PSDest}
\vspace{-15pt}
\end{figure}

The estimated PSD using the PMU data between 1000 to 1500s are shown in Fig. \ref{PSDest}, both of which have relatively blurred spikes, indicating that it is impossible to distinguish the oscillation mechanisms by eyeballing the width of the spike as suggested by \cite{wang2016data}.

\subsection{Case A}

The time series data shown in Fig. \ref{timeseriesA} indicates that the systems evolves from the old steady state (0-300s) to the new steady state (700s-1500s) probably after some change of parameters. The estimated PSD of the new steady state clearly shows that there is an oscillation around 0.12 Hz, yet the PSD has certain bandwidth. We intend to apply the proposed method to figure out the exact mechanism of the oscillation.

\begin{figure}[!ht]
\vspace{-15pt}
\centering
\includegraphics[width=3in,center]{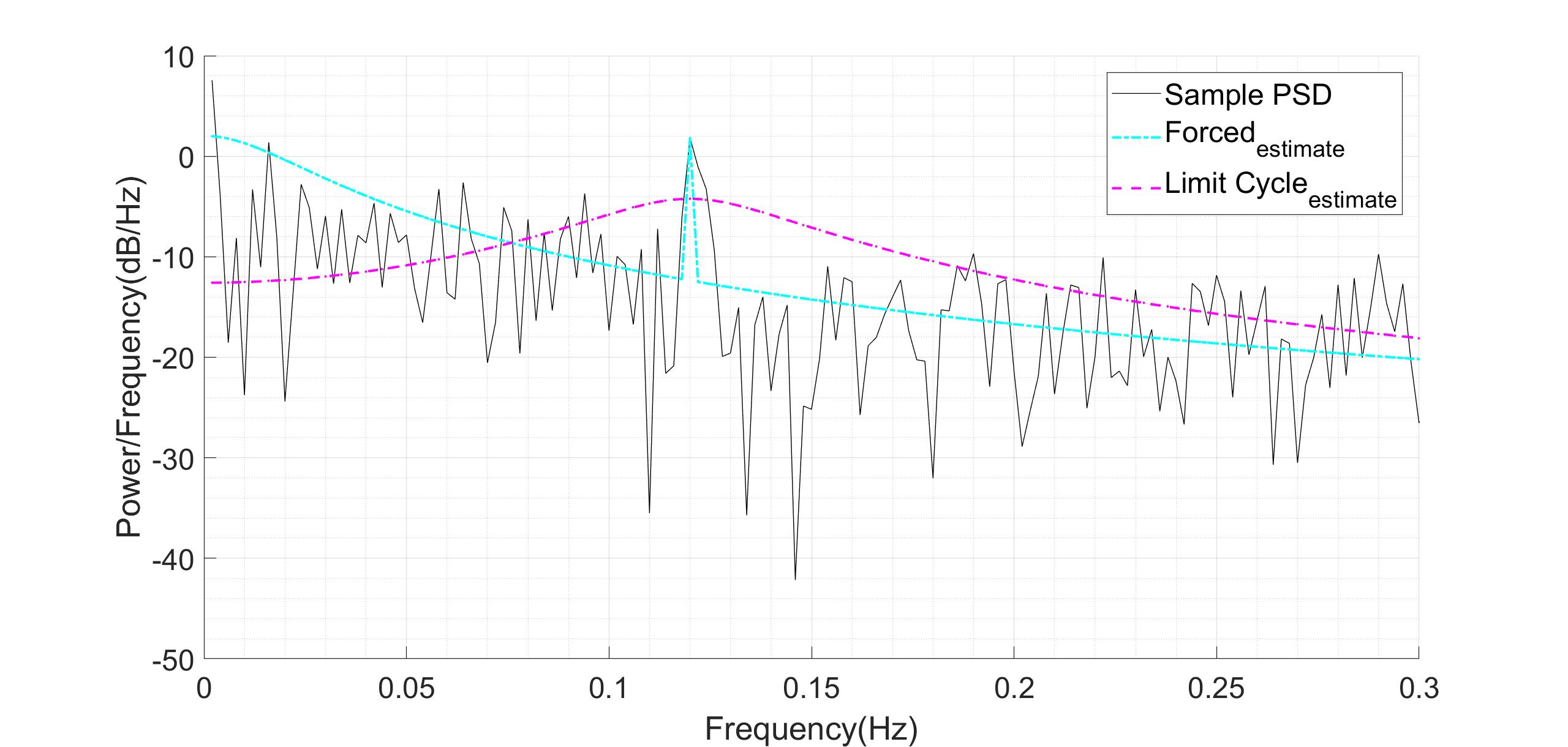}
\caption{The PSD Fitting Result for Case A}\label{PSDfittingA}
\end{figure}

Following the procedure proposed, we get two sets of parameter estimates by solving two constrained optimization problems under different assumptions. The fitted PSDs are presented in Fig. \ref{PSDfittingA}. The parameter values and the loss functions are listed in Table. \ref{parameterestA}. It can be seen the parameter set under the assumption of limit cycle has less loss function value, which is therefore used to compute the classifier $V_{PS}$. Since $V_{PS}=9.82e-16$, much less than $0.5$, the oscillation mechanism is classified to be limit cycle.

\begin{table}[!ht]
\scalebox{0.9}{
\begin{tabular}{ccccccc}
      Constraint & Loss Function & $\beta$ & $\lambda$ & $\alpha$ & $\sigma_1$ & $\sigma_2$  \\
      \hline Limit Cycle &3.99 &1359.11 &0.27 &0.75 &0.15 &1e-05 \\
      \hline Forced Oscillation &6.03 &0.15 & 0.61 &0.75 &1e-05 & 0.18 \\
\end{tabular}}
\caption{Parameter Estimation Result for Case A}\label{parameterestA}
\end{table}

The physical cause of the oscillation has been detailed in \cite{wang2016data}. Starting from 300s, the exponential recovery loads at Bus 5 and Bus 12 increase gradually. By
420s, both loads grow by 12\% and stop increasing afterwards. 
The competing effect between load
dynamics and the over excitation limiters of the generators leads to the voltage instability. The power system passes the Hopf bifurcation point around
400s when a stable limit cycle is born.

\subsection{Case B}
\begin{figure}[!ht]\vspace{-10pt}
\centering
\includegraphics[width=3in,center]{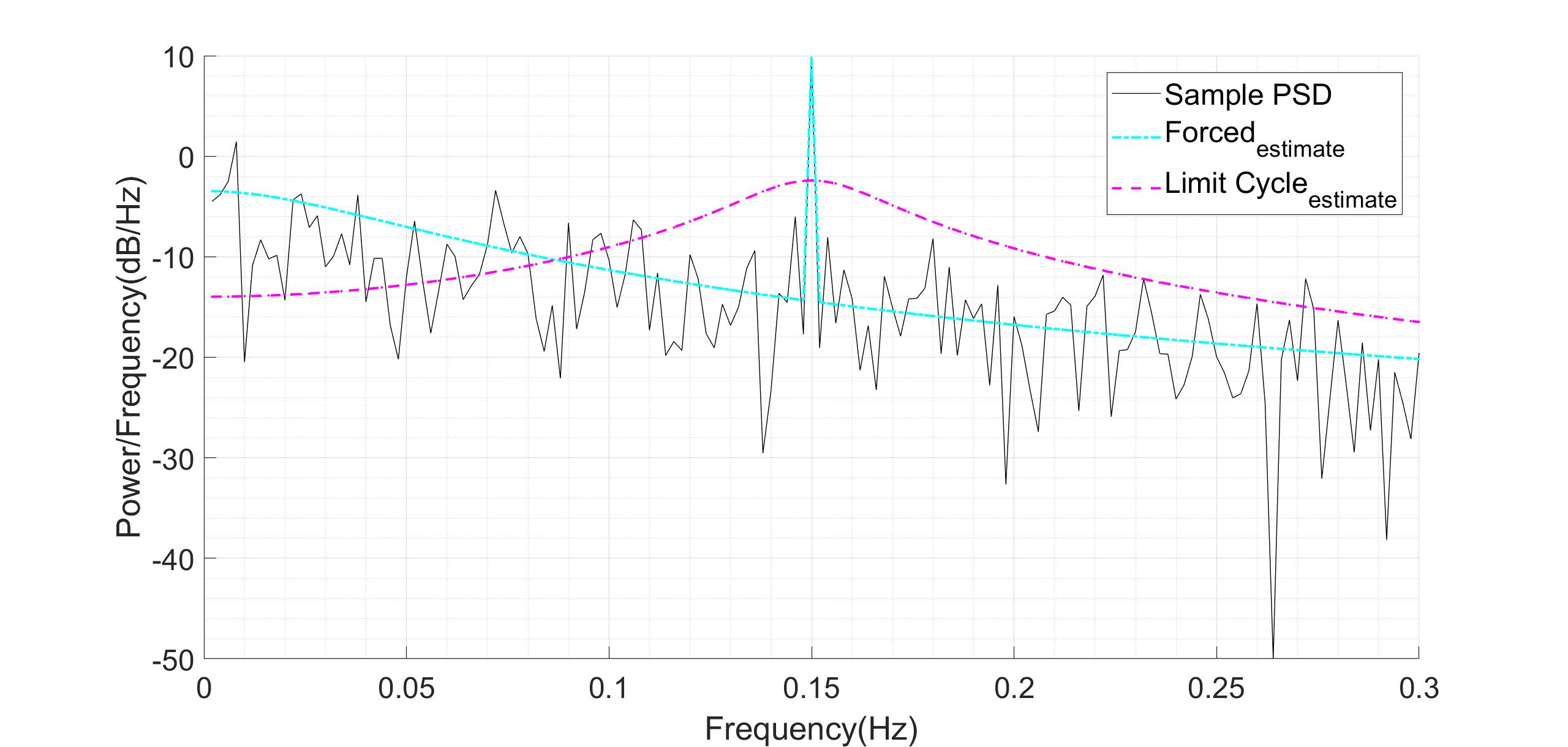}
\caption{The PSD Fitting Result for Case B}\label{PSDfittingB}
\vspace{-10pt}
\end{figure}

In case B, the estimated PSD of the new steady state shown in Fig. \ref{PSDestB} does not have a perfectly thin spike, making the identification based on the bandwidth around the peak \cite{wang2016data} fails. Applying the proposed method, we fit the estimated PSD and obtain two sets of parameters by solving two optimization problems under different assumptions. The results are shown in Fig. \ref{PSDfittingB} and Table. \ref{parameterestB}. 

The loss function of the forced oscillation constraint is smaller,  leading to a classifier value $V_{PS}=1$. Therefore, the oscillation mechanism is identified to be forced oscillation. 

\begin{table}[!ht]
\scalebox{0.9}{
\centering
\begin{tabular}{ccccccc}
      Constraint & Loss Function & $\beta$ & $\lambda$ & $\alpha$ & $\sigma_1$ & $\sigma_2$  \\
      \hline Limit Cycle &4.87 &1318.79 &0.35 &0.94 &0.18 &1e-05 \\
      \hline Forced Oscillation &2.30 &0.28 &0.61 &0.94 &1e-05 &0.19 \\
\end{tabular}}
\caption{Parameter Estimation Result for Case B}\label{parameterestB}
\end{table}

The actual situation is that the exponential recovery
load at Bus 5 increases by 5\% between 300s and 350s, at which one cyclic load joins at Bus 5 with a forced frequency 0.15Hz. Therefore, the oscillation is indeed a forced oscillation. More importantly, although the forced oscillation frequency is very close to the natural frequency mode 0.12Hz, the proposed method is still able to provide accurate identification result. 

\section{Conclusions and Perspectives}
This paper has proposed a periodogram-based method to distinguish natural oscillation and forced oscillation. It has been shown that the limit cycle and the forced oscillation can be described by the same stochastic differential equation yet with different parameter values, based on which a PSD-based classifier is proposed to distinguish different mechanisms. It has been shown via numerical simulation that the proposed method can provide accurate classification results even though the forced oscillation frequency is close to the natural one. Further investigation about locating the source of forced oscillation is needed.

\section{Appendix}
\small
\subsection{Derivation of Equation \ref{eq:forcedsoln}}\label{appendix-1}
\begin{equation}
x(t)=\frac{F}{\sqrt{\gamma^2+(\Omega-\omega_0)^2}}\cos{(\Omega t)}+\sigma\int_0^te^{-\gamma(t-s)}dB_s
\end{equation}
Taking time derivative on both sides we obtain:
\begin{equation}
\dot{x}=-\Omega\frac{F}{\sqrt{\gamma^2+(\Omega-\omega_0)^2}}\sin{(\Omega t)}-\gamma\sigma\int_0^te^{-\gamma(t-s)}dB_s+\sigma\xi(t)    
\end{equation}
Let $\dot{x}=-\gamma(x(t)-Q(t))+\sigma\xi(t)$ and solve for $Q(t)$: 
\begin{align*}
Q(t)&=\frac{\dot{x}(t)+\gamma x(t)-\sigma\xi(t)}{\gamma}\\
&=\frac{F}{\sqrt{\gamma^2+(\Omega-\omega_0)^2}}\cos{(\Omega t)}-\frac{\Omega}{\gamma}\frac{F}{\sqrt{\gamma^2+(\Omega-\omega_0)^2}}\sin{(\Omega t)}\\
&=\frac{F}{\gamma}\sqrt{\frac{\gamma^2+\Omega^2}{\gamma^2+(\Omega-\omega_0)^2}}\sin{(\arctan{\frac{\gamma}{\Omega}} - \Omega t)}
\end{align*}
Note that the phase shift $\arctan{\frac{\gamma}{\Omega}}$ depends on the initial condition. Without loss of generality, $Q(t)$ takes the form:
\begin{equation}
Q(t)=\frac{F}{\gamma}\sqrt{\frac{\gamma^2+\Omega^2}{\gamma^2+(\Omega-\omega_0)^2}}\cos{( \Omega t)} \nonumber
\end{equation}
\normalsize
\vspace{-8pt}
\bibliographystyle{ieeetran}
\bibliography{references}
\end{document}